\documentclass[preprint,12pt]{elsarticle}
\usepackage{tikz}
\usetikzlibrary{shapes.geometric, arrows, positioning}
\usetikzlibrary{arrows.meta, shadows.blur}
\usepackage{hyperref}
\usepackage{booktabs}
\usepackage{tabularx}
\usepackage{amssymb}
\usepackage{amsmath}
\journal{Renewable Energy}
\usepackage{lineno}
\usepackage{graphicx}
\usepackage{array}
\usepackage{soul}
\usepackage{xcolor}

\begin{document}

\begin{frontmatter}

\title{Multilayer GNN for Predictive Maintenance and Clustering in Power Grids}

\author[ndsu]{Muhammad Kazim\corref{cor1}}
\ead{muhammad.kazim@ndsu.edu}

\author[ndsu]{Harun Pirim\corref{cor2}}
\ead{harun.pirim@ndsu.edu}

\author[chau]{Chau Le}
\author[trung]{Trung Le}
\author[om]{Om Prakash Yadav}

\cortext[cor1]{Corresponding author}
\cortext[cor2]{Corresponding author}

\affiliation[ndsu] {organization={Industrial \& Manufacturing Engineering, North Dakota State University}, 
            addressline={NDSU Dept. 2485, PO Box 6050}, 
            city={Fargo},
            postcode={58108-6050}, 
            state={ND},
            country={USA}}

\affiliation[chau] {organization={Department of Engineering Technology \& Construction Management, University of North Carolina at Charlotte}, 
            postcode={28223}, 
            state={NC},
            country={USA}}

\affiliation[om] {organization={Department of Industrial \& Systems Engineering, North Carolina Agriculture \& Tech State University}, 
            city={Greensboro},
            postcode={27411}, 
            state={NC},
            country={USA}}

\affiliation[trung] {organization={Department of Industrial \& Management Systems Engineering, University of South Florida}, 
            addressline={4202 East Fowler Avenue, ENG 030}, 
            city={Tampa},
            postcode={33620}, 
            state={FL},
            country={USA}}

\begin{abstract}
Unplanned power outages impose significant economic costs, exceeding \$150 billion annually in the U.S., driven by the limitations of current predictive maintenance (PdM) models that overlook spatial, temporal, and causal interdependencies in power grid failures. This study proposes a multilayer Graph Neural Network (GNN) framework to enhance PdM and enable resilience-based substation clustering, leveraging seven years of incident data from Oklahoma Gas \& Electric (292{,}830 records across 347 substations). Our model integrates specialized GNN layers---Graph Attention Networks for spatial dependencies, Graph Convolutional Networks for temporal patterns, and Graph Isomorphism Networks for causal interactions---fused through attention-weighted embeddings. For PdM, the framework achieves a 30-day F1-score of $0.8935 \pm 0.0258$, outperforming industry benchmarks like XGBoost by 3.2\% and Random Forest by 2.7\%, and surpassing single-layer GNN variants by 10--15\%. Ablation studies underscore the causal layer’s importance, with its removal dropping the F1-score to $0.7354 \pm 0.0418$. For resilience analysis, HierarchicalRiskGNN embeddings, clustered via HDBSCAN, identify eight distinct operational risk groups, with the highest-risk cluster (Cluster 5, 44 substations) showing an incident rate of 388.4/year and recovery time of 602.6 minutes, compared to low-risk clusters (e.g., Cluster 2) with rates below 62/year. Statistical analysis (ANOVA, $p < 0.0001$) confirms robust inter-cluster separation, while our clustering outperforms K-Means and Spectral Clustering with a Silhouette Score of 0.626 (vs. 0.315) and a Davies-Bouldin index of 0.527 (vs. 0.806). This approach advances proactive grid management, offering superior predictive accuracy and resilience insights, with future applications in multiclass failure prediction and grid analytics, supporting sustainable energy infrastructure modernization.

\end{abstract}


\end{frontmatter}


\section{Introduction}
\label{sec:intro}

Energy infrastructure forms the backbone of economic productivity and societal functioning. In the United States alone, unplanned power outages cost over \$150 billion annually \citep{biden2023}. To mitigate such disruptions, predictive maintenance (PdM) and resilience-based clustering have emerged as essential tools for proactive grid management. PdM enables the early detection of equipment degradation, reducing unplanned downtime by up to 70\% and lowering maintenance costs by 30\% \citep{doe2010}. Clustering, in parallel, supports operational resilience through zoning strategies such as islanding and prioritized maintenance planning. These techniques, when integrated effectively, provide not only economic benefits but also enhanced reliability and response capability. Predictive analytics in energy infrastructure has demonstrated return-on-investment (ROI) ratios as high as 10:1 \citep{biden2023}, yet existing frameworks fall short of leveraging the full complexity of interdependent power systems.

A core limitation of conventional PdM and clustering methods lies in their treatment of infrastructure assets as isolated units. Traditional models—whether rule-based, statistical, or even machine learning-based—often fail to capture the cascading nature of failures that propagate through interconnected grid components \citep{geng2022predictive, schlapfer2008stress, wu2016modeling, jyoti2023topological}. These approaches neglect spatial proximity, temporal degradation trends, and causal dependencies among assets, resulting in inaccurate forecasts and fragmented resilience strategies \citep{cheng2023identification}. Even graph-based methods, which introduce topological awareness, typically rely on single-layer representations that inadequately model the heterogeneous relationships inherent in modern power grids \citep{jyoti2023topological}.

Previous studies have explored a variety of graph neural network (GNN) architectures to model complex relational data. For instance, spatial-temporal GNNs have been widely applied to dynamic graph problems such as traffic forecasting and social network analysis \citep{yu2018spatio, wu2019graph}. These models effectively capture spatial dependencies between nodes and temporal dynamics over time, making them suitable for systems with evolving interactions. However, in domains like power grids, where relationships between substations may shift due to operational changes or failures, spatial-temporal GNNs often assume a fixed graph structure, limiting their adaptability. This motivates our approach, which incorporates multiple types of dependencies—spatial, temporal, and statistical co-failure relationships—to better model the resilience of power grid substations.

To address these structural limitations, multilayer network theory provides a formal framework for modeling interdependent infrastructures by preserving both intra-layer and inter-layer dynamics \citep{domenico_mathematical_2013}. In the context of power systems, nodes such as substations, transformers, and transmission lines often participate in diverse relational layers—ranging from physical proximity to temporal fault sequences and causal pathways. Multilayer modeling captures these relationships explicitly, overcoming the loss of information typical in aggregated graphs \citep{nicosia2015measuring, puxeddu2021, kazim2025link}. Applications of multilayer analysis in the energy sector have demonstrated its value in identifying critical nodes whose failures induce cascading disruptions \citep{nosyrev2019}, and in characterizing the resilience of coupled gas-electricity systems \citep{kazim2024}. These findings establish multilayer modeling as a powerful tool for understanding and mitigating systemic vulnerabilities in energy infrastructures.

Building on this foundation, Graph Neural Networks (GNNs) have emerged as a scalable and data-driven approach for learning from graph-structured data in PdM and resilience tasks. In power grids, GNNs have been used for overload prediction \citep{zhang2024graph}, equipment health indexing \citep{ifeanyi2024graph}, and cyber-physical interaction modeling \citep{Islam}. However, most existing GNN-based PdM models in energy systems focus exclusively on spatial features and ignore the temporal and causal aspects critical to forecasting real-world failures. Moreover, these models generally employ single-layer graphs, missing the nuanced, layered interdependencies across system components \citep{xia2022}. As a result, they fail to capture how failures evolve, how causal chains propagate, and how risks concentrate across regions and subsystems.

Multilayer GNNs offer a promising solution by integrating spatial, temporal, and failure propagation patterns into a unified predictive framework. Early multilayer GNN approaches, such as supra-graph modeling \citep{Grassia2022}, simplify the representation but overlook the semantics of individual layers. While multilayer GNNs like MGCN \citep{ghorbani2019mgcn} and attention-based multi-relational GNNs \citep{meng2024deepmcgcn} have advanced layer-wise information fusion, they are primarily designed for entity-type heterogeneity rather than the spatiotemporal-causal dynamics essential for power grid modeling. For instance, MGCN focuses on integrating multiple relation types but does not explicitly account for temporal evolution or causal propagation of failures. Similarly, \citet{li2023type} proposed a multi-channel GNN for event detection in power systems but overlooked explicit multilayer graph structures and relied on static type labels, limiting its adaptability to dynamic grid conditions. \citet{zhang2024interpretable} demonstrated that multilayer GNNs can outperform single-layer variants by 10–15\% in wind power forecasting. However, a cohesive, end-to-end framework for multilayer PdM in power systems—one that combines topology, time-series behavior, and failure causality—remains lacking. Unlike prior studies that apply multilayer GNNs to isolated tasks, such as wind power forecasting \citep{zhang2024interpretable}, our approach uniquely unifies predictive maintenance and resilience clustering within a single scalable framework, providing both predictive accuracy and actionable operational insights for grid management.

Clustering in multilayer power grids plays a critical role in resilience planning, particularly for segmenting infrastructure into operational zones based on vulnerability or failure patterns. Traditional clustering methods—such as K-Means, Spectral Clustering, or even single-layer Louvain methods—fail to incorporate inter-layer dependencies and often assume equal importance across all features. This fragmentation limits the ability to systematically prioritize risk and segment the grid in a way that aligns with both predictive maintenance outcomes and real-time operational constraints. While generalized multilayer clustering techniques \citep{huang2021survey} address some of these issues by optimizing inter-layer modularity, they seldom integrate predictive layers (e.g., causal fault propagation) with operational relevance. In power systems, multilayer clustering has been applied for controlled islanding—partitioning the grid into self-sustaining zones during disruptions \citep{amini2023segmentation} and for maintenance planning \citep{losapio2024state}, but these approaches often treat layers heuristically or independently. Moreover, traditional methods like K-Means or Spectral Clustering are often inadequate for power grid applications. K-Means relies on Euclidean distances, which fail to account for the non-linear, relational dependencies between substations, such as those driven by electrical flows or temporal failure patterns. Similarly, Spectral Clustering, which leverages graph Laplacians, assumes a static graph structure and may overlook dynamic or multi-relational interactions, such as co-failures over time \citep{amini2023segmentation}. These gaps necessitate a more sophisticated approach to resilience grouping that captures the complex interplay of spatial, temporal, and statistical dependencies in power grids.

Despite a growing body of work applying multilayer GNNs to individual energy tasks \cite{li2023type, zhang2024graph, zhao2024interpretable}, no existing framework unifies multilayer predictive maintenance and resilience-based clustering within a scalable architecture tailored for power grids.

To bridge this gap, our study makes three key contributions: (1) We propose a multilayer GNN-based predictive maintenance framework that captures spatial, temporal, and causal interdependencies using layer-specific message passing, enabling the model to learn tailored representations for each edge modality based on its distinct relational semantics; (2) We introduce a risk-aware clustering approach that incorporates exogenous failure drivers (e.g., vegetation, lightning, weather) into graph embeddings for infrastructure segmentation and prioritization; and (3) We validate our approach on real-world power grid data, demonstrating its effectiveness in identifying high-risk substations, forecasting failures, and supporting operationally meaningful resilience clustering (i.e., groupings that reflect actual outage patterns, recovery times, and maintenance priorities). Crucially, the architecture is designed for scalability and modularity: it supports large-scale, heterogeneous grid data and enables multi-task learning through modular layer-wise encoders, shared representations across tasks, and efficient graph sampling. By unifying multilayer predictive maintenance and risk-based clustering into a single framework, this work advances a proactive, interpretable, and scalable paradigm for energy system management.

The remainder of this paper is organized as follows. Section 2 introduces the proposed multilayer GNN model and computational framework. Section 3 presents experimental evaluations and baseline comparisons. Section 4 concludes with practical implications and future research directions.

\section{Methodology}
\label{sec:method}
\begin{figure}[htbp]
    \centering
    \includegraphics[width=0.95\textwidth]{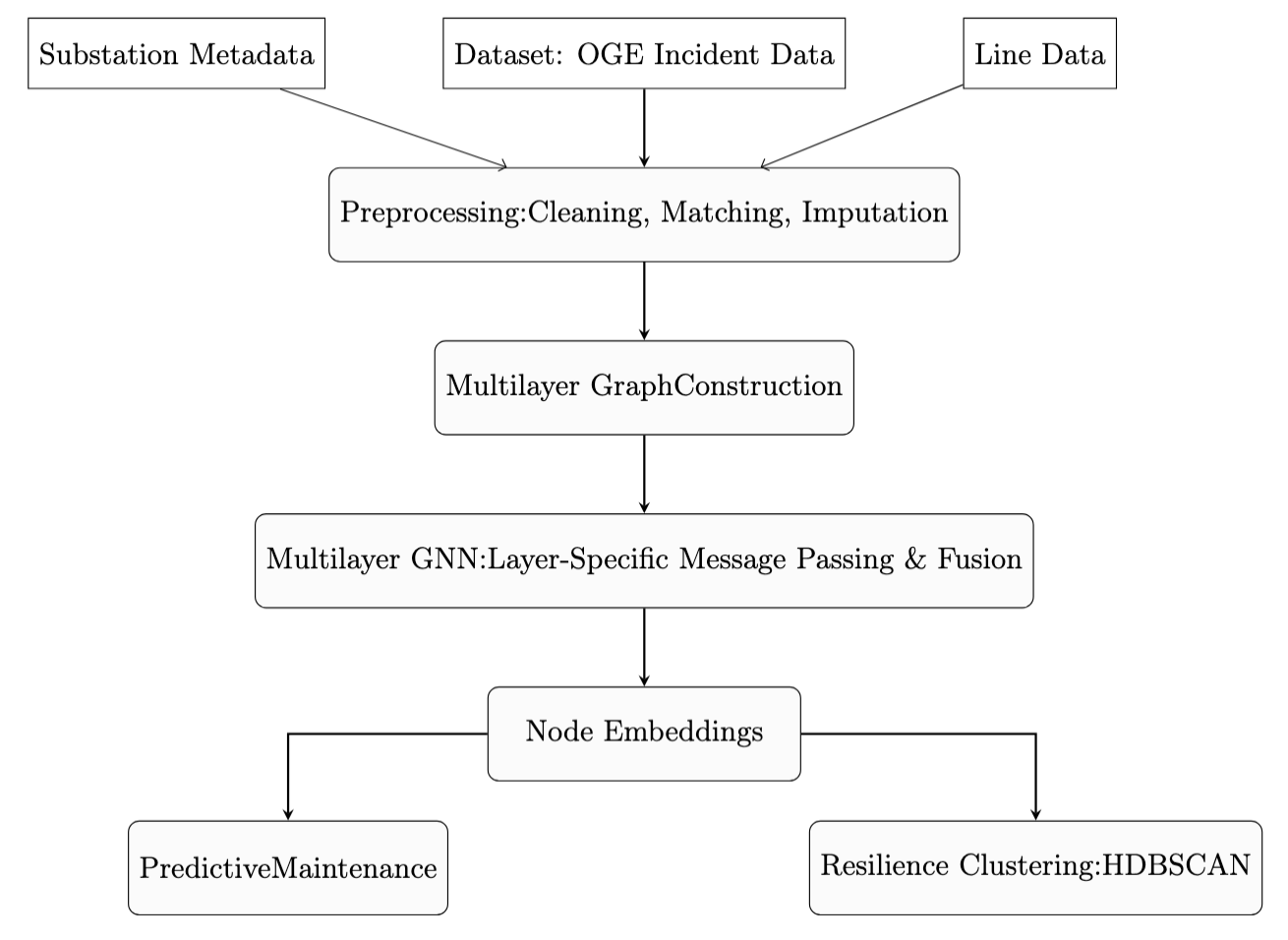}
\caption{Methodology pipeline. Each block represents a major stage in the framework—from raw data preprocessing to multilayer graph construction, GNN-based embedding, and downstream tasks (predictive maintenance and clustering).}
    \label{fig:full-method-flow}
\end{figure}

\begin{figure}[htbp]
    \centering
    \includegraphics[width=0.95\textwidth]{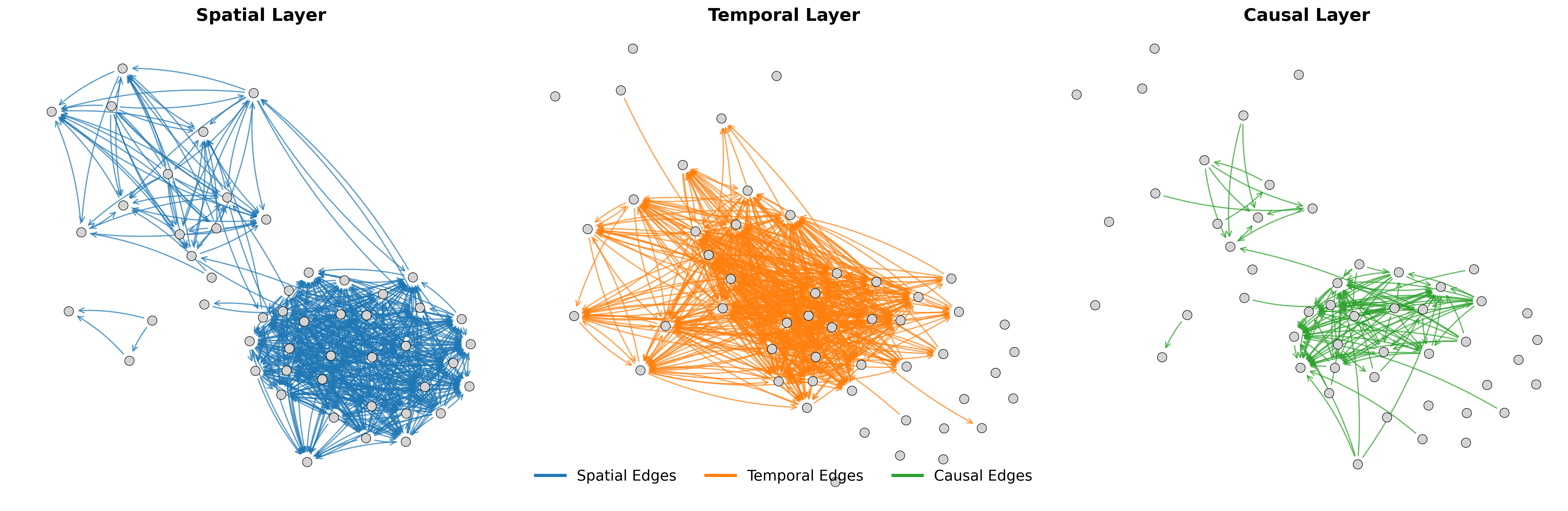}
    \caption{Multilayer network architecture used in graph construction. Nodes represent substations; edges encode spatial (physical/geographic), temporal (event-driven), and causal (statistical co-occurrence) dependencies. This corresponds to the “Multilayer Graph Construction” block in Figure~\ref{fig:full-method-flow}.}
    \label{fig:network}
\end{figure}

Building upon the motivations outlined in Section~\ref{sec:intro}, our methodology integrates a multilayer graph representation of the Oklahoma Gas \& Electric (OGE) power grid with specialized Graph Neural Network (GNN) architectures to address two primary objectives: predictive maintenance (PdM) and resilience-based clustering. As illustrated in Figure~\ref{fig:full-method-flow}, the overall pipeline proceeds from data preprocessing to multilayer graph construction, followed by GNN-based embedding and downstream tasks.  To capture the complex interdependencies that underlie power infrastructure failures, we construct three distinct edge layers—spatial, temporal, and causal—(Fig.~\ref{fig:network}) that reflect physical, co-occurrence, and inferred failure relationships, respectively, which corresponds to the “Multilayer Graph Construction” block in the flowchart. Together, these figures provide an overview of both the system pipeline and the underlying graph topology used for learning.  This section details the dataset characteristics, preprocessing pipeline, multilayer graph construction, PdM target definition, GNN architecture, and the subsequent resilience clustering strategy.

\subsection{Dataset Characteristics}
\label{subsec:data_char}

We utilize the OGE incident dataset spanning 2015--2021, comprising 292{,}830 records across 347 substations. The dataset provides seven years of operational data, averaging approximately 4.2 incidents per day. Each record includes 52 attributes, including event timestamps, failure causes, equipment types, and feeder details. Substations are geospatially distributed, with coordinates recorded at a latitude-longitude precision of $10^{-6}$, enabling fine-grained spatial modeling.

\begin{figure}[htbp]
    \centering
    \includegraphics[width=0.6\textwidth]{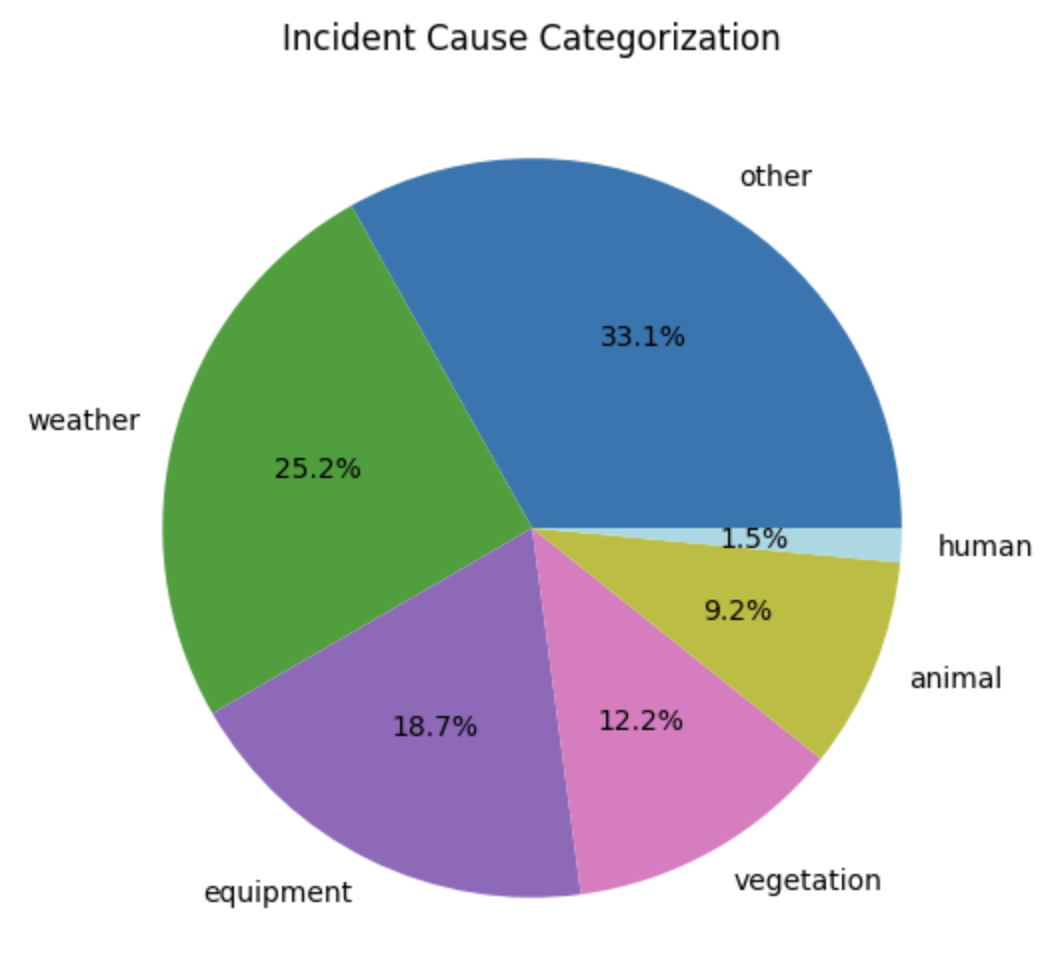}
    \caption{Incident cause distribution showing weather-related predominance}
    \label{fig:causes}
\end{figure}

Figure~\ref{fig:causes} illustrates the predominance of weather-related failures (25.2\%) and equipment malfunctions (18.7\%), underscoring both the diversity and the imbalance across failure types. While weather incidents dominate, vegetation-related and equipment-induced failures remain operationally significant. These characteristics highlight the complexity of the dataset and motivate the need for a modeling approach that captures multiple relational dimensions—temporal, spatial, and causal.

\subsection{Preprocessing Pipeline}
\label{subsec:preprocessing}

The raw dataset exhibits inconsistencies across several dimensions, including substation identifiers, missing timestamps, and unaligned metadata. We implement a multi-stage preprocessing pipeline to ensure data integrity and consistency.

\subsubsection{Data Cleaning}
We remove or impute records lacking valid timestamps (ensuring $t_{\text{off}} \leq t_{\text{on}}$) and exclude incidents whose substations cannot be reliably matched. The cleaned dataset is defined as:

\[
\mathcal{D}_{\text{clean}} = \{d \in \mathcal{D}_{\text{raw}} \mid \text{valid\_times}(d) \,\wedge\, \text{has\_substation}(d)\},
\]
where $\text{valid\_times}(d)$ enforces temporal consistency and $\text{has\_substation}(d)$ verifies the presence of a valid substation ID.

\subsubsection{Geospatial Reconciliation}
Substation names in the incident logs often deviate from standardized identifiers in the reference dataset (\texttt{Substation.csv}). We standardize identifiers (e.g., \texttt{FEEDER:NAME}) and apply fuzzy matching using Levenshtein distance~\cite{levenshtein1966binary} with a threshold $\tau \approx 0.85$. Records with unresolved mismatches are flagged for manual review or excluded.

\subsubsection{Voltage Imputation and Metadata}
We enrich substation metadata by imputing missing voltage levels via:
\begin{enumerate}
    \item \emph{String-based Extraction:} Parsing voltage tags from line descriptions (e.g., \texttt{69kV}).
    \item \emph{Feeder-Based Voting:} Computing the mode voltage among associated feeders.
    \item \emph{Regional Defaults:} Assigning the most common voltage in the geographic vicinity.
\end{enumerate}
This process ensures consistent voltage classification across the network.

\subsection{Multilayer Network Construction}
\label{subsec:network_construction}

We construct a weighted multilayer heterogeneous graph $\mathcal{G}$ to represent inter-substation dependencies across spatial, temporal, and causal dimensions. The graph is defined as:
\[
\mathcal{G} = \bigl(\mathcal{V}, \mathcal{E}_{\text{spatial}}, \mathcal{E}_{\text{temporal}}, \mathcal{E}_{\text{causal}}\bigr),
\]
where each node $v \in \mathcal{V}$ corresponds to a substation, and each weighted edge layer encodes a distinct class of relational structure.

\textbf{Node Features ($\mathcal{V}$):}
The graph contains $|\mathcal{V}| = 347$ substations, each represented by an 8-dimensional numerical feature vector comprising:

\textbf{Node Features ($\mathcal{V}$):}
Each of the $|\mathcal{V}| = 347$ substations is represented by an 8-dimensional numeric vector comprising:

\begin{itemize}
    \item[i)] \textit{Geospatial}: Latitude, Longitude
    \item[ii)] \textit{Topological}: Number of connections, Avg. line voltage, Line length (km)
    \item[iii)] \textit{Incident Statistics}: Historical incident count, Mean SAIFI
    \item[iv)] \textit{Electrical}: Nominal voltage
    \item[v)] \textit{Risk Indicators}: Vegetation, Lightning, Weather, and Equipment failure counts
\end{itemize}

Categorical metadata (e.g., substation type via \texttt{PLANTCLASS}) is separately one-hot encoded and appended. All numeric features are standardized using \texttt{StandardScaler}, except geospatial coordinates, which are preserved in raw form to maintain spatial fidelity.

\textbf{Edge Layers ($\mathcal{E}$):}
All three edge layers are weighted, with edge-specific attributes incorporated into the downstream GNN. 
Specifically: 

\begin{itemize}
    \item $\mathcal{E}_{\text{spatial}}$: Encodes physical transmission lines and proximity-based substation relationships. include either a unit weight for physical connections or a distance-decay weight for proximity-based links (see Section~\ref{subsubsec:spatial_layer}).
    
    \item $\mathcal{E}_{\text{temporal}}$: Captures time-decayed weights based on inter-incident time gaps and co-occurrence of outage events across substations (see Section~\ref{subsubsec:temporal_layer}).
    
    \item $\mathcal{E}_{\text{causal}}$: Represents statistically enriched correlations such as standardized co-occurrence scores and frequency ratios in outage causes (e.g., lightning, vegetation)  between substations (see Section~\ref{subsubsec:causal_layer}).
\end{itemize}

These weights inform the layer-specific message-passing operations during GNN training and enhance the model’s capacity to differentiate structural roles across modalities.

\textbf{Graph Assembly:}
We integrate node features and edge layers into a unified heterogeneous graph using the PyTorch Geometric \texttt{HeteroData} class. Each substation is indexed via a unique identifier, and edge types are explicitly annotated as triplets of the form:
\[
(\texttt{substation}, \; \texttt{edge\_type}, \; \texttt{substation})
\]
where \texttt{edge\_type} $\in$ \{\texttt{spatial}, \texttt{temporal}, \texttt{causal}\}.

Edge attributes are type-specific: spatial and temporal edges include physical and proximity metadata (e.g., line length, shared feeders), while causal edges incorporate co-occurrence statistics and one-hot encoded cause labels. The final heterogeneous graph object is stored in PyTorch's binary \texttt{.pt} format using \texttt{torch.save()} for efficient loading in downstream graph learning workflows.

\subsubsection{Spatial Layer ($\mathcal{E}_{\text{spatial}}$):} 
\label{subsubsec:spatial_layer}

This layer comprises 23,378 edges constructed from two complementary sources of connectivity: physical transmission infrastructure and geographic proximity.

\textbf{(1) Physical Transmission Lines:} 
We extract physical connectivity using \texttt{Lines.csv}, which specifies named transmission lines and their associated voltage classes (\texttt{Const\_Volt}). For each line, substation names are parsed, cleaned, and matched using fuzzy string similarity. Only matches with a confidence score $\geq$ 64 are automatically retained, while those between 55 and 64 are flagged for manual review. For every valid pair of substations connected via a transmission line, we record:
\begin{itemize}
    \item \texttt{has\_line} = 1 (indicates physical infrastructure),
    \item \texttt{line\_voltage} (taken from the original line data),
    \item \texttt{line\_length\_km} (computed from geospatial shape length).
\end{itemize}

\textbf{(2) Geographic Proximity:} 
We identify additional edges based on spatial closeness. Specifically, we compute the pairwise Haversine distances (in kilometers) between substations that were co-affected by the same weather-related events. The proximity threshold is set to the 75th percentile of these distances, calculated from 1,203 co-occurring weather incidents, to ensure relevance to shared environmental exposure. Substation pairs within this threshold and sharing the same voltage class are connected with:

\begin{itemize}
    \item \texttt{is\_nearby} = 1 (indicates spatial adjacency),
    \item \texttt{line\_voltage} = null (not derived from physical transmission infrastructure),
    \item \texttt{line\_length\_km} = 0 (no actual line; proximity-only connection).
\end{itemize}

\textbf{(3) Shared Context Features:} 
For each edge, whether physical or proximity-based, we compute:
\begin{itemize}
    \item \texttt{shared\_cities}: count of overlapping job cities affected by incidents,
    \item \texttt{shared\_feeders}: count of feeders shared between connected substations.
\end{itemize}

\textbf{(4) Distance-Based Weighting:} 
Each edge includes a geodesic distance attribute (\texttt{distance\_km}), and a distance-decay weight is assigned as:
\[
\texttt{weight} = 
\begin{cases}
1.0 & \text{if has\_line = 1 (physical connection)} \\
\frac{1}{1 + \texttt{distance\_km}} & \text{if is\_nearby = 1 (proximity-based)}
\end{cases}
\]

This formulation ensures that the spatial graph encodes both explicit physical infrastructure and latent proximity-based relationships relevant to regional stressors (e.g., storms, wildfires). The resulting edge set $\mathcal{E}_{\text{spatial}}$ serves as the structural foundation for spatial reasoning in downstream GNN layers.

\subsubsection{Temporal Layer ($\mathcal{E}_{\text{temporal}}$)}
\label{subsubsec:temporal_layer}

The temporal layer encodes short-term causal or co-occurrence relationships between substations based on incident timestamps. We define an adaptive yet globally fixed time window $\theta$ for constructing temporal edges by calculating the 80th percentile of inter-incident arrival times across the entire dataset. Specifically, incident timestamps are sorted, and the difference between consecutive outage events is computed to form a distribution of inter-arrival intervals. The 80th percentile of this distribution yields a data-driven temporal window $\theta$ that captures typical co-occurrence dynamics while avoiding excessive noise. This window is uniformly applied across all substations and outage types, making it globally adaptive but not localized per node. The chosen percentile offers a balance—lower thresholds risk omitting valid correlations, while higher values can lead to overly dense edge sets. While we do not vary the percentile in this study, the chosen 80th percentile was informed by empirical inspection of outage dynamics and was found to produce a well-balanced temporal edge network. This parameter remains configurable, allowing utilities or researchers to tailor the temporal sensitivity of the model to different grid environments or operational contexts:

\[
\theta = \text{percentile}_{80}\left(\texttt{Job OFF Time}_{i+1} - \texttt{Job OFF Time}_i\right), \quad \text{for } i = 1, \ldots, N-1
\]

\textbf{Edge Construction:} For every incident at substation $u$, we identify subsequent incidents at substation $v$ that occur within $\theta$ minutes. Each such temporal pair $(u, v)$ is assigned a weight:
\[
w_{uv}^t = \exp\left(-\frac{|t_u - t_v|}{60}\right)
\]
This exponential decay (with a 60-minute decay constant) reflects the intuition that the influence or shared causal likelihood of events decreases over time.

\textbf{Co-Occurrence Filtering:} After aggregating temporal edges by source-target pairs, we apply an automatic thresholding strategy to filter out spurious or low-frequency pairs. Specifically, we retain edges that co-occur at least $k$ times, where:
\[
k = \max\left(3, \text{median}\left(\text{co\_occurrence\_count}_{uv}\right)\right)
\]

(The lower bound of 3 is empirically chosen to suppress noise from random or isolated events while ensuring statistical relevance in sparse regions.)

This conservative rule ensures that only frequently co-occurring substation pairs are retained in the final temporal edge set.

The resulting temporal graph $\mathcal{E}_{\text{temporal}}$ captures meaningful patterns of temporally correlated outages—often shaped by storm clusters, operational cascades, or weather-driven propagation.

\subsubsection{Causal Layer ($\mathcal{E}_{\text{causal}}$)}
\label{subsubsec:causal_layer}
This layer captures statistically significant failure correlations between substations, conditioned on both causal attribution and spatiotemporal proximity. It is important to note that the term ``causal layer'' refers to empirically enriched co-failure correlations rather than formal causal discovery. Specifically, this layer identifies statistically significant patterns of co-occurring failures, which may indicate shared vulnerabilities or propagation pathways, but does not establish true causal relationships. This approach is inspired by methods in ecology and network science that use co-occurrence data to infer dependencies~\cite{araujo2011using}, and it complements traditional causal inference techniques~\cite{pearl2009causality} by providing a data-driven approximation of interdependence tailored to the predictive maintenance context.

\textbf{(1) Cause-Specific Temporal Windows:}
For each failure cause \( c \in \mathcal{C} \), we compute an adaptive time window \( \theta_c \) as:

\[
\theta_c = \min\left(\text{percentile}_{75}\left(\Delta t_i^{(c)}\right), \theta_{\text{max}}\right), \quad \forall\, c \in \mathcal{C}
\]

This formulation computes the 75\textsuperscript{th} percentile of inter-arrival times \( \Delta t_i^{(c)} \) between incidents attributed to each cause \( c \), allowing the temporal window to adapt to the underlying frequency of different failure types—whether bursty (e.g., lightning) or gradual (e.g., deterioration).

\textbf{(2) Spatially-Constrained Matching:}
To ensure physical plausibility, co-occurrence analysis is only performed between substation pairs $(u, v)$ that are directly connected in the spatial layer $\mathcal{E}_{\text{spatial}}$.

\textbf{(3) Statistical Enrichment:}
For each substation pair $(u, v)$ and failure cause $c$, we count the number of temporal co-occurrences $\text{obs}_{uv}^c$ within $\theta_c$ hours. We then estimate the expected number of co-occurrences under an independence assumption:
\[
\mathbb{E}\left[\text{obs}_{uv}^c\right] = \lambda_u \cdot \lambda_v \cdot \theta_c \cdot T
\]
where $\lambda_u$ and $\lambda_v$ denote failure rates at $u$ and $v$, and $T$ is the observation period in days.

Assuming Poisson-distributed co-occurrences, we compute a standardized \( z \)-score to assess statistical enrichment:

\[
Z_{uv}^c = \frac{\text{obs}_{uv}^c - \mathbb{E}[\text{obs}_{uv}^c]}{\sqrt{\mathbb{E}[\text{obs}_{uv}^c]}}
\]

This enrichment step highlights pairs of substations that experience significantly more co-occurring failures than expected by chance, indicating potential causal links or shared vulnerabilities.

\textbf{(4) Pruning and Feature Extraction:}
Edges are retained if $Z_{uv}^c$ exceeds the 85th percentile of the empirical distribution, and $\text{obs}_{uv}^c \ge 3$. For each retained edge, we record:
\begin{itemize}
    \item \texttt{cause}: the shared failure cause $c$,
    \item \texttt{z\_score}: standardized co-occurrence signal,
    \item \texttt{cooccur\_ratio}: empirical-to-expected ratio,
    \item \texttt{window\_hrs}: adaptive window used,
    \item \texttt{cooccur\_count}: number of observed co-occurrences.
\end{itemize}

This layer reveals hidden propagation patterns in the grid—e.g., repeated vegetation failures along shared corridors or widespread lightning damage in regional clusters—by combining domain knowledge, spatiotemporal alignment, and statistical filtering.

\subsection{Predictive Maintenance Target}
\label{subsec:targets_PdM}

We define a binary substation-level target label $y_i \in \{0,1\}$ to indicate whether a substation requires proactive replacement or major maintenance. Based on the filtered incident data, we apply the following procedure:

\begin{enumerate}
    \item \emph{Identify ``Severe'' Causes:} We classify outage causes as severe if they exceed the 90\textsuperscript{th} percentile in both outage duration and the maximum number of customers affected.
    
    \item \emph{Tag Critical Equipment:} We identify equipment categories as critical if they exhibit a high frequency of failure and an average SAIDI (System Average Interruption Duration Index) above the 90\textsuperscript{th} percentile.
    
    \item \emph{Check Maintenance Window:} A substation is labeled $y_i = 1$ if it experiences a severe-cause outage involving critical equipment and does not encounter another major incident within 180 days. Otherwise, $y_i = 0$.
\end{enumerate}

The labeling condition can be expressed formally as:

\[
y_i = \mathbb{I}\left[\text{SAIDI}_i > Q_{90}(\text{SAIDI}) \land \Delta t_{\text{next}} > 180 \text{ days}\right],
\]

where $\text{SAIDI}_i$ denotes the historical System Average Interruption Duration Index for substation $i$ (with the 90\textsuperscript{th} percentile threshold equal to 8.7 hours), and $\Delta t_{\text{next}}$ represents the time until the next severe failure at the same substation. This approach targets high-impact outages while ensuring a sufficient window for proactive maintenance. Applying this criterion results in 19.4\% positive labels (67 out of 347 substations). To address class imbalance during model training, we employ focal loss~\cite{lin2017focal}, which down-weights easy examples and focuses learning on hard-to-classify cases.

\subsection{Multi-Layer GNN Architecture}
\label{subsec:model}

To generate predictive embeddings for each substation and classify the \emph{needs\_replacement} label ($y_i$), we design a specialized Graph Neural Network (GNN) that processes each relational layer independently and then fuses the resulting embeddings. The architecture comprises two main components:

\subsubsection{Layer-Specific Message Passing}

We employ separate GNN modules for each edge type—\textbf{spatial}, \textbf{temporal}, and \textbf{causal}—to enable targeted message passing tailored to the unique semantics and statistical properties of each relational layer. This separation allows the model to learn modality-specific representations, as spatial edges reflect physical proximity, temporal edges encode time-sensitive co-occurrence patterns, and causal edges capture enriched statistical dependencies. Each node's features are normalized with respect to the edge type, and edge attributes (e.g., distance, temporal decay, causal strength) are incorporated into the respective message-passing functions:

\begin{itemize}
    \item \textbf{Spatial Layer}: A Graph Attention Network (GATv2)~\cite{brody2022gatv2} models physical and geographic interactions by leveraging edge features such as voltage level and physical distance. GATv2 dynamically assigns attention weights to neighbors, enabling the model to prioritize influential spatial connections.
    
    \item \textbf{Temporal Layer}: A GATv2 convolutional layer~\cite{brody2022gatv2} encodes event co-occurrence within a learned time window, using exponential decay weights to modulate temporal relevance. This formulation allows the model to capture short-term temporal patterns in failure propagation.
    
    \item \textbf{Causal Layer}: A Graph Isomorphism Network (GIN)~\cite{xu2019powerful} models complex causal dependencies between failures. GIN is chosen for its expressive capability to differentiate graph structures, which is essential for capturing intricate cause-effect relationships in outage propagation.
\end{itemize}

Each edge type is processed using two stacked GNN layers with graph normalization and dropout to stabilize learning. The outputs are layer-specific embeddings capturing structural and relational information unique to each connection type.

\subsubsection{Attention-Based Fusion and Prediction}

To combine the layer-specific embeddings, we apply an attention-based fusion mechanism that learns to weight each embedding via trainable attention scores:

\[
\mathbf{h}_i^{\text{(fused)}} = \Phi\left(\alpha_s \cdot \mathbf{h}_i^{\text{(spatial)}} \,\oplus\, \alpha_t \cdot \mathbf{h}_i^{\text{(temporal)}} \,\oplus\, \alpha_c \cdot \mathbf{h}_i^{\text{(causal)}}\right),
\]

where $\oplus$ denotes vector concatenation, and $\Phi(\cdot)$ represents a feed-forward neural network (typically composed of one or two multilayer perceptron layers). The attention weights $\alpha_s$, $\alpha_t$, and $\alpha_c$ are normalized coefficients learned via multi-head self-attention, enabling the model to adaptively weigh the contribution of each relational modality during fusion. The resulting fused embedding $\mathbf{h}_i^{\text{(fused)}}$ is passed through a final linear layer followed by a log-softmax activation to produce a two-class probability distribution.

\subsubsection{Training and Evaluation}

The model is trained using the \textit{AdamW} optimizer~\cite{loshchilov2017decoupled} with a learning rate of $10^{-3}$ and weight decay of $10^{-5}$. Gradient clipping is applied with an $\ell_2$-norm threshold of $||\nabla||_2 \leq 1.0$ to stabilize training. We employ early stopping based on validation loss to prevent overfitting.

To address the class imbalance in the target labels, we use the focal loss function~\cite{lin2017focal} with parameters $\alpha = 0.75$ and $\gamma = 2$. This formulation emphasizes hard-to-classify positive samples by down-weighting easy negatives. The focal loss is defined as:

\[
\mathcal{L}_\text{focal} = -\,\alpha \,(1 - p_t)^\gamma \,\log(p_t),
\]

where $p_t$ is the predicted probability of the true class, and $\alpha$ and $\gamma$ control the degree of focus on misclassified examples. The optimizer is coupled with a learning rate scheduler that reduces the learning rate on validation loss plateaus, further enhancing convergence stability.

The dataset is partitioned using a temporal split to mimic real-world deployment settings: data from 2015–2018 is used for training, 2019 for validation, and 2020–2021 for testing. For robustness, we perform three-fold cross-validation on each time window (30, 60, and 180 days), and report averaged metrics across folds.

\subsection{Resilience Clustering}
\label{subsec:clustering}
To support long-term resilience planning, we perform clustering of substations in the learned embedding space produced by a hierarchical GNN, where spatial, temporal, and causal relations are encoded through layer-specific modules and integrated via attention-based fusion. These clusters represent groups of substations with similar spatial, causal, and temporal vulnerability profiles.

\subsubsection{Hierarchical GNN for Risk-Aware Embedding}
We adopt a \emph{HierarchicalRiskGNN} that generates node embeddings by encoding information from all three edge types:

\[
\mathbf{z}_i = \text{Attention}\left(
    \text{GATv2}_{\text{spatial}}(\mathbf{x}_i),\,
    \text{GATv2}_{\text{temporal}}(\mathbf{x}_i),\,
    \text{GIN}_{\text{causal}}(\mathbf{x}_i)
\right)
\]

Each encoder operates on a specific relational layer, and the final embedding $\mathbf{z}_i$ is obtained via multi-head attention across these modalities. Training is guided by a multi-objective loss comprising:

\begin{itemize}
    \item \textbf{Risk Prediction Loss:} Predicts vegetation, weather, lightning, and equipment risk scores using MSE loss against historical incident frequencies.
    \item \textbf{Topological Consistency Loss:} Encourages similar embeddings for directly connected nodes via cosine similarity objectives.
    \item \textbf{Cluster Separation Loss:} Encourages entropy-based separation and intra-cluster compactness using a deep clustering head.
\end{itemize}

The embeddings are normalized and optimized over 200 epochs using the AdamW optimizer with cyclical learning rates and early stopping based on validation loss.

\subsubsection{Clustering in Latent Space}
We apply UMAP~\cite{mcinnes2018umap} for dimensionality reduction (to $\mathbb{R}^3$) followed by HDBSCAN~\cite{campello2013density} clustering on the embedding space. HDBSCAN infers the number of clusters and marks outliers as noise. We set:
\begin{itemize}
    \item \texttt{min\_cluster\_size} = 15
    \item \texttt{min\_samples} = 5
    \item \texttt{cluster\_selection\_epsilon} = 0.5
\end{itemize}

These settings yield robust, interpretable clusters that group substations with similar multi-risk exposure patterns.

\subsubsection{Cluster Interpretation and Evaluation}

To support resilience planning, we interpret the learned clusters through a structured post-processing pipeline. Once embeddings are generated from the Hierarchical GNN, we apply dimensionality reduction (e.g., UMAP) to map high-dimensional embeddings to a lower-dimensional space while preserving local structure. This facilitates downstream analysis and visualization.

Cluster assignments are obtained using HDBSCAN ~\cite{campello2013density}, which automatically infers the number of clusters and identifies noise points without requiring a fixed cluster count. To evaluate the structure and separability of clusters, we employ the following methods:

\begin{itemize}
    \item \textbf{Risk Profiling:} Each cluster's average values across risk features (e.g., vegetation, lightning, weather, equipment) are aggregated to characterize dominant threat types within each group.
    
    \item \textbf{Topological Cohesion:} Cosine similarity between nodes and their neighbors is used to verify that the learned embeddings preserve meaningful relationships across spatial, temporal, and causal layers.
    
    \item \textbf{Reliability Trends:} Survival analysis methods (e.g., Kaplan–Meier estimation~\cite{kaplan1958nonparametric}) are applied using outage frequency as a proxy for failure likelihood, enabling comparative analysis across cluster profiles.
    
    \item \textbf{Geospatial Distribution:} Cluster labels are projected onto the physical grid layout to examine spatial coherence and uncover geographically consistent regions affected by similar risks.
\end{itemize}

These steps do not depend on ground truth labels but instead rely on an unsupervised structure emerging from the graph. This allows utilities to interpret resilience clusters in terms of risk heterogeneity, maintenance needs, and regional vulnerabilities, providing a foundation for downstream planning and decision-making.

\subsection{Baseline Models and Comparative Framework}
\label{subsec:baseline_models}

To contextualize the performance of our graph-based learning pipeline, we establish two baseline strategies tailored to each downstream task.

\paragraph{(1) Predictive Maintenance (PdM):}
We implement classical machine learning models using node-level features extracted from the heterogeneous graph. Specifically, we train:

\begin{itemize}
    \item \textbf{Random Forest:} Using 100 trees and class-balanced weighting~\cite{breiman2001random}.
    \item \textbf{XGBoost:} With depth-3 trees and scale-sensitive weighting based on class imbalance~\cite{chen2016xgboost}.
\end{itemize}

To ensure temporal robustness, target labels and features are constructed fold-wise: for each evaluation fold, incident records beyond the latest training timestamp are excluded from both feature computation and label assignment. This design avoids future leakage and mirrors realistic deployment. Features are drawn directly from the substation node representations (including risk scores), and targets are computed using the same maintenance criterion defined in Section~\ref{subsec:targets_PdM}.

\paragraph{(2) Resilience Clustering.}
We compare our GNN-based embeddings against traditional clustering methods applied to raw node features. In particular:

\begin{itemize}
    \item \textbf{K-Means Clustering:} Performed on standardized node features with $k \in \{2,3,4\}$, selecting the best configuration via silhouette score~\cite{rousseeuw1987silhouettes}.
    \item \textbf{Spectral Clustering:} Using a precomputed spatial adjacency matrix as affinity, also evaluated across $k \in \{2,3,4\}$~\cite{ng2002spectral}.
\end{itemize}

All baselines are evaluated using the same downstream metrics—silhouette score~\cite{rousseeuw1987silhouettes}, Davies–Bouldin index, and ANOVA-based risk variance—to ensure consistent and fair benchmarking.

This comparative framework enables isolation of the gains attributable to GNN-driven representation learning versus traditional feature engineering or unsupervised clustering alone.

In the subsequent section, we present experimental results illustrating how our approach outperforms traditional single-layer baselines and provides valuable operational insights for energy utilities.

\section{Results and Discussion}
\label{sec:results}

This section presents the experimental outcomes of our multilayer Graph Neural Network (GNN) framework applied to the Oklahoma Gas \& Electric (OGE) dataset, as introduced in Section~\ref{sec:intro} and elaborated in Section~\ref{sec:method}. We evaluate the model's performance on two core tasks: \textbf{predictive maintenance (PdM)} and \textbf{resilience clustering}. For PdM, we assess the model's ability to predict substation maintenance needs across 30, 60, and 180-day windows, leveraging the spatial, temporal, and causal layers defined in Section~\ref{subsec:network_construction}. For resilience clustering, we analyze the quality and operational utility of substation clusters derived from GNN embeddings (Section~\ref{subsec:clustering}). We compare our multilayer GNN against baseline methods to demonstrate its effectiveness in capturing heterogeneous interdependencies, and we discuss the implications for energy system management, alongside limitations and future research directions.

\subsection{Predictive Maintenance Performance}
\label{subsec:pdm_results}

The PdM task is formulated as a binary classification problem, predicting whether a substation requires proactive replacement or major maintenance ($y_i = 1$) or not ($y_i = 0$), as outlined in Section~\ref{subsec:targets_PdM}. Given the class imbalance (19.4\% positive labels), we report accuracy, precision, recall, and F1-score to provide a comprehensive evaluation, with focal loss employed during training to address this imbalance (Section~\ref{subsec:model}).

\subsubsection{Quantitative Results for Different Time Windows}
\label{subsubsec:pdm_quantitative}

We assessed PdM performance across three maintenance planning horizons—30, 60, and 180 days—using the multilayer GNN with stratified 3-fold cross-validation. For each fold, the model was trained on a subset of substations and tested on the remaining substations, with targets generated independently for each fold using only historical data up to the training cutoff, ensuring no temporal leakage. The dataset spans 2015–2021, with stratified 3-fold cross-validation across substations. Training ran for up to 100 epochs. The average test set performance metrics across folds are presented in Tables~\ref{tab:pdm_180}, \ref{tab:pdm_60}, and \ref{tab:pdm_30}.

\paragraph{180-Day Window}
For the 180-day window, the multilayer GNN achieved an accuracy of $0.8444 \pm 0.0241$, precision of $0.8225 \pm 0.0456$, recall of $0.8763 \pm 0.0261$, and F1-score of $0.8472 \pm 0.0179$. These results indicate robust identification of substations needing maintenance over a longer horizon, balancing false positives and missed detections effectively.

\begin{table}[htbp]
    \centering
    \caption{Predictive Maintenance Performance for 180-Day Window}
    \label{tab:pdm_180}
    \begin{tabularx}{\linewidth}{lXXXX}
        \toprule
        \textbf{Model} & \textbf{Accuracy} & \textbf{Precision} & \textbf{Recall} & \textbf{F1-score} \\
        \midrule
        Multi\_GNN & $0.8444 \pm 0.0241$ & $0.8225 \pm 0.0456$ & $0.8763 \pm 0.0261$ & $0.8472 \pm 0.0179$ \\
        SPATIAL\_ONLY & $0.7232 \pm 0.0293$ & $0.6883 \pm 0.0155$ & $0.7934 \pm 0.0815$ & $0.7354 \pm 0.0418$ \\
        TEMPORAL\_ONLY & $0.8156 \pm 0.0159$ & $0.7921 \pm 0.0234$ & $0.8473 \pm 0.0322$ & $0.8182 \pm 0.0157$ \\
        CAUSAL\_ONLY & $0.8387 \pm 0.0211$ & $0.8281 \pm 0.0248$ & $0.8471 \pm 0.0160$ & $0.8374 \pm 0.0198$ \\
        Random Forest & $0.8473 \pm 0.0383$ & $0.8530 \pm 0.0732$ & $0.8412 \pm 0.0235$ & $0.8450 \pm 0.0328$ \\
        XGBoost & $0.8155 \pm 0.0254$ & $0.8038 \pm 0.0394$ & $0.8294 \pm 0.0432$ & $0.8150 \pm 0.0228$ \\
        \bottomrule
    \end{tabularx}
\end{table}

\paragraph{60-Day Window}
For the 60-day window, performance improved, with an accuracy of $0.8588 \pm 0.0148$, precision of $0.8817 \pm 0.0152$, recall of $0.8586 \pm 0.0128$, and F1-score of $0.8700 \pm 0.0136$. The higher precision suggests fewer false positives, advantageous for resource-constrained maintenance planning over mid-term horizons.

\begin{table}[htbp]
    \centering
    \caption{Predictive Maintenance Performance for 60-Day Window}
    \label{tab:pdm_60}
    \begin{tabularx}{\linewidth}{lXXXX}        \toprule
        \textbf{Model} & \textbf{Accuracy} & \textbf{Precision} & \textbf{Recall} & \textbf{F1-score} \\
        \midrule
        Multi\_GNN & $0.8588 \pm 0.0148$ & $0.8817 \pm 0.0152$ & $0.8586 \pm 0.0128$ & $0.8700 \pm 0.0136$ \\
        SPATIAL\_ONLY & $0.6916 \pm 0.0249$ & $0.7188 \pm 0.0221$ & $0.7226 \pm 0.0359$ & $0.7204 \pm 0.0254$ \\
        TEMPORAL\_ONLY & $0.8502 \pm 0.0141$ & $0.8649 \pm 0.0249$ & $0.8638 \pm 0.0079$ & $0.8641 \pm 0.0105$ \\
        CAUSAL\_ONLY & $0.8472 \pm 0.0229$ & $0.8842 \pm 0.0335$ & $0.8324 \pm 0.0081$ & $0.8573 \pm 0.0198$ \\
        Random Forest & $0.8531 \pm 0.0374$ & $0.8820 \pm 0.0481$ & $0.8482 \pm 0.0306$ & $0.8642 \pm 0.0336$ \\
        XGBoost & $0.8299 \pm 0.0179$ & $0.8635 \pm 0.0235$ & $0.8220 \pm 0.0348$ & $0.8415 \pm 0.0178$ \\
        \bottomrule
    \end{tabularx}
\end{table}

\paragraph{30-Day Window}
The 30-day window yielded the highest performance: accuracy of $0.8846 \pm 0.0253$, precision of $0.9336 \pm 0.0131$, recall of $0.8576 \pm 0.0412$, and F1-score of $0.8935 \pm 0.0258$. The elevated precision reflects the model's strength in pinpointing urgent maintenance needs with minimal false positives, ideal for short-term operational decisions.

\begin{table}[htbp]
    \centering
    \caption{Predictive Maintenance Performance for 30-Day Window}
    \label{tab:pdm_30}
    \begin{tabularx}{\linewidth}{lXXXX}        \toprule
        \textbf{Model} & \textbf{Accuracy} & \textbf{Precision} & \textbf{Recall} & \textbf{F1-score} \\
        \midrule
        Multi\_GNN & $0.8846 \pm 0.0253$ & $0.9336 \pm 0.0131$ & $0.8576 \pm 0.0412$ & $0.8935 \pm 0.0258$ \\
        SPATIAL\_ONLY & $0.7319 \pm 0.0319$ & $0.7602 \pm 0.0276$ & $0.7713 \pm 0.0513$ & $0.7649 \pm 0.0328$ \\
        TEMPORAL\_ONLY & $0.8674 \pm 0.0216$ & $0.9077 \pm 0.0410$ & $0.8575 \pm 0.0572$ & $0.8795 \pm 0.0219$ \\
        CAUSAL\_ONLY & $0.8673 \pm 0.0324$ & $0.9081 \pm 0.0024$ & $0.8524 \pm 0.0638$ & $0.8781 \pm 0.0345$ \\
        Random Forest & $0.8531 \pm 0.0134$ & $0.8945 \pm 0.0314$ & $0.8427 \pm 0.0295$ & $0.8669 \pm 0.0111$ \\
        XGBoost & $0.8500 \pm 0.0257$ & $0.8842 \pm 0.0337$ & $0.8476 \pm 0.0173$ & $0.8653 \pm 0.0224$ \\
        \bottomrule
    \end{tabularx}
\end{table}

\subsubsection{Comparison with Baselines}
\label{subsubsec:pdm_baselines}

To validate the multilayer GNN's integration of spatial, temporal, and causal layers, we compared it against:
\begin{itemize}
    \item \textbf{Single-layer GNNs}: Models using only one edge type (\texttt{SPATIAL\_ONLY}, \texttt{TEMPORAL\_ONLY}, \texttt{CAUSAL\_ONLY}).
    \item \textbf{Non-graph Methods}: Random Forest and XGBoost trained on flattened feature vectors.
\end{itemize}

Across all time windows, the multilayer GNN (\texttt{Multi\_GNN}) outperformed baselines, particularly in F1-score. For the 180-day window, it achieved an F1-score of $0.8472$, surpassing \texttt{SPATIAL\_ONLY} ($0.7354$) by 15\% and Random Forest ($0.8450$) slightly. Similar trends held for 60-day and 30-day windows, with the largest gains over single-layer models (10–15\%), echoing improvements noted in wind power forecasting (Section~\ref{sec:intro}). This underscores the value of modeling heterogeneous interdependencies jointly.

\subsubsection{Ablation Studies}
\label{subsubsec:pdm_ablation}

To isolate the individual contribution of each relational modality, we conduct targeted ablation studies by training the model using only one edge layer at a time—removing the remaining two. Unlike the broader benchmarking in Table~\ref{tab:pdm_180}, which compares full models and classical baselines, this analysis focuses solely on the internal structure of the multilayer GNN to understand the marginal utility of each layer in isolation. Results for the 180-day window are presented in Table~\ref{tab:ablation_180}, with similar trends observed across shorter horizons.

\begin{table}[htbp]
    \centering
    \caption{Ablation Study for 180-Day Window (Single-Layer Performance)}
    \label{tab:ablation_180}
    \begin{tabularx}{\linewidth}{lXXXX}
        \toprule
        \textbf{Configuration} & \textbf{Accuracy} & \textbf{Precision} & \textbf{Recall} & \textbf{F1-score} \\
        \midrule
        \textbf{Full Model} & \textbf{0.8444 ± 0.0241} & \textbf{0.8225 ± 0.0456} & \textbf{0.8763 ± 0.0261} & \textbf{0.8472 ± 0.0179} \\
        Only Spatial Layer & $0.7232 \pm 0.0293$ & $0.6883 \pm 0.0155$ & $0.7934 \pm 0.0815$ & $0.7354 \pm 0.0418$ \\
        Only Temporal Layer & $0.8156 \pm 0.0159$ & $0.7921 \pm 0.0234$ & $0.8473 \pm 0.0322$ & $0.8182 \pm 0.0157$ \\
        Only Causal Layer & \textbf{0.8387} ± \textbf{0.0211} & \textbf{0.8281} ± \textbf{0.0248} & \textbf{0.8471} ± \textbf{0.0160} & \textbf{0.8374} ± \textbf{0.0198} \\

        \bottomrule
    \end{tabularx}
\end{table}

The greatest performance degradation occurs when using only the spatial layer (F1-score: $0.7354$), highlighting its limited standalone predictive power. In contrast, the causal layer alone achieves an F1-score of $0.8374$, closely matching the full model. This underscores the causal layer's high discriminative strength and validates the design choice to explicitly model fault propagation mechanisms (Section~\ref{sec:intro}).

\subsection{Resilience-Oriented Substation Clustering}
\label{subsec:clustering_results}

Using the HierarchicalRiskGNN framework, we generated substation-level embeddings that integrate spatial, temporal, and causal interdependencies across the Oklahoma Gas \& Electric (OGE) network. Each edge modality—spatial, temporal, and causal—is processed through a dedicated GNN encoder, with the resulting representations fused via multi-head self-attention. This architecture allows the model to learn the relative importance of each relational modality when shaping the final node embeddings.

The clustering pipeline involved three stages: (1) representation learning via \textit{HierarchicalRiskGNN}, (2) nonlinear manifold reduction using UMAP with cosine similarity, and (3) density-based clustering using HDBSCAN (\texttt{min\_cluster\_size}=15, \texttt{epsilon}=0.25), which is well-suited for identifying variable-density structures in high-dimensional spaces~\cite{campello2013density}.

This process yielded \textbf{eight distinct operational clusters}, with a Silhouette Score of 0.626 and a Davies-Bouldin Index of 0.527, indicating well-separated and internally coherent groupings. Out of 347 substations, 328 (94.5\%) were assigned to clusters, while 19 (5.5\%) were designated as outliers due to insufficient local density and excluded from aggregate cluster analyses (see Table~\ref{tab:cluster_stats}).

\begin{table}[htbp]
\centering
\caption{Cluster Membership Summary}
\label{tab:cluster_stats}
\begin{tabular}{lcccc}
\toprule
\textbf{Cluster ID} & \textbf{n\_substations} & \textbf{\% of Total} & \textbf{Incidents/yr} & \textbf{Recovery (min)} \\
\midrule
0 & 55 & 16.8\% & 63.0  & 567.4 \\
1 & 37 & 11.3\% & 63.7  & 392.1 \\
2 & 43 & 13.1\% & 61.9  & 333.8 \\
3 & 24 & 7.3\%  & 126.9 & 521.5 \\
4 & 31 & 9.5\%  & 205.5 & 343.9 \\
5 & 44 & 13.4\% & 388.4 & 602.6 \\
6 & 36 & 11.0\% & 84.5  & 260.1 \\
7 & 58 & 17.7\% & 153.1 & 358.4 \\
\bottomrule
\end{tabular}
\end{table}

This segmentation reveals operationally meaningful patterns. Notably, Cluster 5 exhibits the highest outage frequency and recovery time, indicating a high-risk zone, while Clusters 0 through 2 consistently fall on the low-impact spectrum. These results underscore the model’s ability to generate interpretable embeddings and support proactive, risk-informed infrastructure management.

\begin{figure}[htbp]
    \centering
    \includegraphics[width=0.95\textwidth]{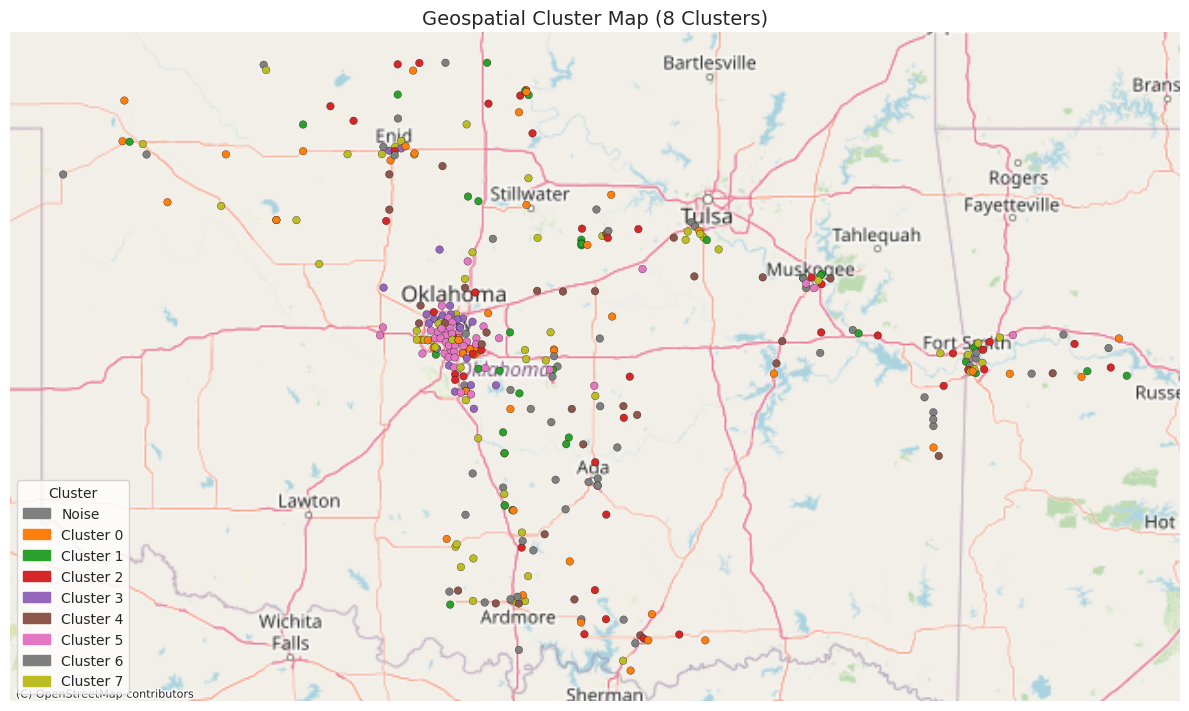}
    \caption{Geospatial Distribution of Substation Clusters (HDBSCAN, 8 Clusters)}
    \label{fig:geospatial}
\end{figure}

Figure~\ref{fig:geospatial} illustrates the spatial dispersion of clusters across Oklahoma and surrounding regions. Notably, high-risk clusters (e.g., Cluster 5 and Cluster 4) are concentrated near densely interconnected urban and industrial corridors, while low-risk clusters such as Cluster 0 and Cluster 1 are more spatially dispersed. This suggests that geographical or environmental exposure has an influence on risk stratification.

\subsubsection{Structural Validation}
\label{subsubsec:structure_validation}
To evaluate the topological coherence of clusters, we analyzed intra-cluster edge densities across the heterogeneous network. The graph includes spatial, temporal, and causal edges—each capturing distinct dimensions of substation relationships.

\begin{itemize}
    \item \textbf{Spatial edges:} 6,965 of 23,378 (29.8\% intra-cluster)
    \item \textbf{Causal edges:} 4,382 of 10,136 (43.2\% intra-cluster)
    \item \textbf{Temporal edges:} 13,992 of 37,924 (36.9\% intra-cluster)
\end{itemize}

While some edge types naturally span clusters (e.g., temporal), the observed intra-cluster ratios exceed random expectation and validate that the model learns meaningful structural patterns. Notably, causal edges exhibit the strongest intra-cluster alignment, consistent with the idea that shared failure mechanisms are more predictive of operational similarity.

\subsubsection{Risk Profile Differentiation}
\label{subsubsec:risk_profiles}

To assess whether the clusters reflect meaningful differences in risk, we evaluated the distribution of four normalized risk indicators: Vegetation, Lightning, Weather, and Equipment. ANOVA tests across all clusters yielded statistically significant differentiation:

\begin{itemize}
    \item Vegetation: $F = 131.32$, $p = 0.0001$
    \item Lightning: $F = 86.38$, $p = 0.0001$
    \item Weather: $F = 182.78$, $p = 0.0001$
    \item Equipment: $F = 148.05$, $p = 0.0001$
\end{itemize}

Here, $p = 0.0001$ indicates extremely strong statistical significance, confirming that the observed differences in risk exposure across clusters are unlikely to have occurred by chance. Clusters 5 and 4 consistently score highest on risk exposure, whereas Clusters 0, 1, and 2 form the low-risk tier. This validates that the model's embeddings preserve and enhance risk information in the clustering space. To enhance interpretability, we conducted a post-hoc analysis of the dominant features driving membership in the high-risk clusters. Cluster 5, for example, exhibited higher-than-average values for historical incident frequency, vegetation-related outages, and lightning density. These patterns were not only consistent across multiple clustering runs but also spatially aligned with urban-industrial corridors exposed to severe weather and vegetation overgrowth. Substations in this cluster also had elevated causal propagation scores, indicating their potential to both initiate and receive cascading failures. These factors provide a domain-consistent explanation for the high-risk label and demonstrate that the GNN embedding preserves interpretable and operationally relevant signals.

\subsubsection{Operational Impact and Reliability Metrics}
\label{subsubsec:operational_reliability}

To ground the clustering results in real-world operational performance, we computed standard substation-level reliability metrics using raw outage records, including:

\begin{itemize}
    \item Mean incidents per year
    \item Average recovery time (minutes)
    \item Mean customer minutes interrupted (CMI)
\end{itemize}

\begin{figure}[htbp]
    \centering
    \includegraphics[width=\textwidth]{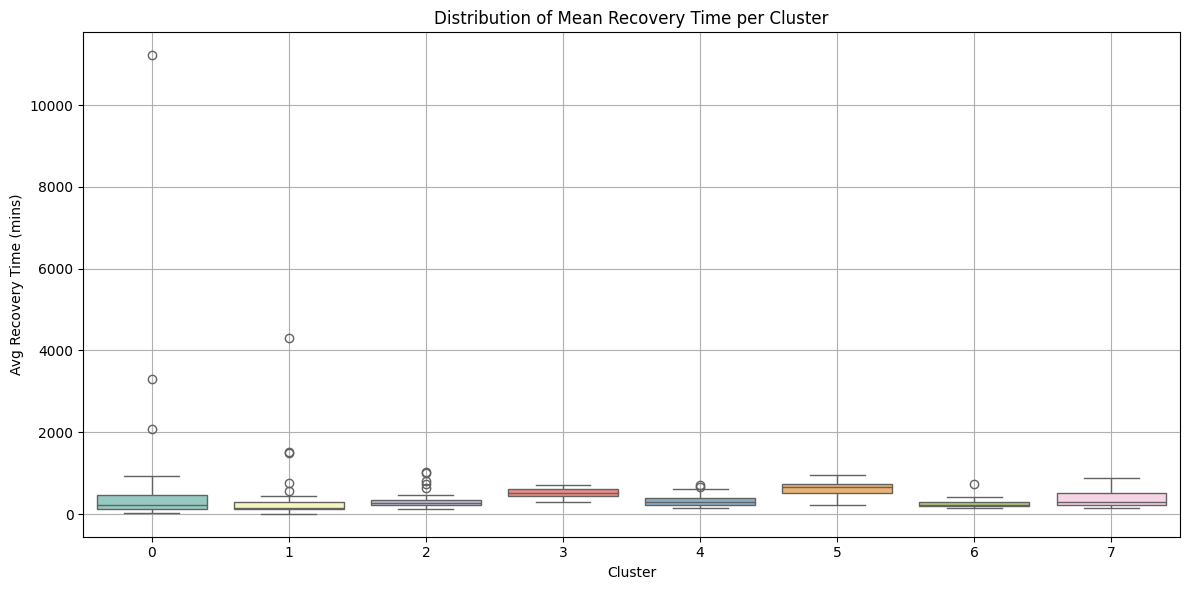}
    \caption{Distribution of Mean Recovery Time by Cluster. Clusters 5 and 3 exhibit the longest average outage durations, while Clusters 6, 1, and 0 show consistently fast recoveries, indicating greater resilience.}
    \label{fig:recovery_box}
\end{figure}

\begin{figure}[htbp]
    \centering
    \includegraphics[width=0.7\textwidth]{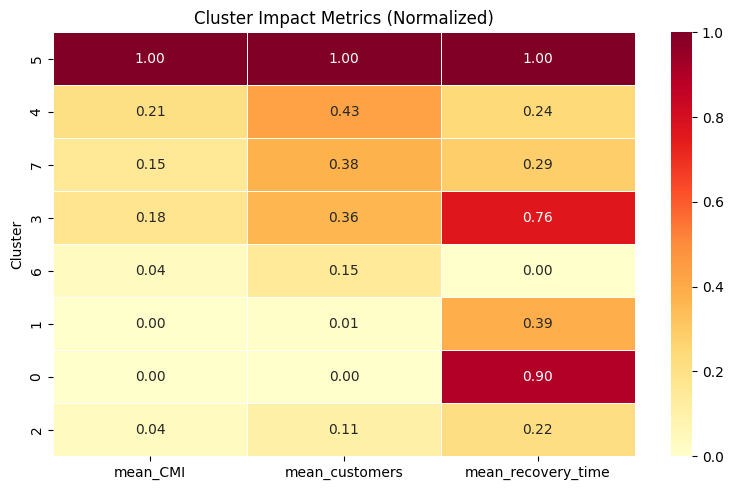}
    \caption{Cluster Impact Heatmap (Normalized). Cluster 5 stands out as the most operationally critical across all three metrics: customer minutes interrupted (CMI), customers affected, and recovery time.}
    \label{fig:impact_heatmap}
\end{figure}

\noindent Cluster 5 is the most operationally critical, with the highest average CMI (107.6M), the longest recovery time (602.6 minutes), and the highest incident frequency. These patterns confirm that cluster membership corresponds to observable infrastructure performance.

\subsubsection{Comparative Benchmarking}
\label{subsubsec:clustering_benchmarks}

We benchmarked our clustering approach against traditional methods—KMeans and Spectral Clustering—using the same UMAP-reduced embeddings for consistency. Table~\ref{tab:clustering_bench} summarizes the evaluation using the Silhouette score (for cluster cohesion/separation), the Davies-Bouldin index (for intra-cluster similarity), and the average ANOVA $p$-value across all risk factors (for risk discrimination power):

\begin{table}[htbp]
    \centering
    \caption{Benchmarking Cluster Quality Across Methods}
    \label{tab:clustering_bench}
    \begin{tabular}{lccc}
        \toprule
        \textbf{Method} & \textbf{Silhouette} & \textbf{Davies-Bouldin} & \textbf{Avg Risk $p$-value} \\
        \midrule
        \textbf{Our GNN (HDBSCAN)} & \textbf{0.626} & \textbf{0.527} & \textbf{0.0001} \\
        KMeans                     & 0.315          & 0.806          & 0.0001 \\
        Spectral Clustering        & -0.004         & 8.994          & 0.0715 \\
        \bottomrule
    \end{tabular}
\end{table}

Our multilayer GNN-based clustering substantially outperforms the baselines:

\begin{itemize}
    \item \textbf{Cluster Cohesion:} A Silhouette score of $0.626$ reflects strong internal cohesion and well-separated clusters—nearly double that of KMeans ($0.315$) and far superior to Spectral Clustering, which failed to produce meaningful groupings (score = $-0.004$).
    
    \item \textbf{Cluster Compactness:} The Davies-Bouldin index ($0.527$) is significantly lower than KMeans ($0.806$) and dramatically better than Spectral Clustering ($8.994$), indicating compact and distinct cluster boundaries.

    \item \textbf{Risk Differentiation:} ANOVA yields an average $p$-value of $0.0000$ for our GNN model and KMeans, suggesting strong differentiation across risk dimensions. However, our method achieves this with far superior structural clustering quality, unlike K-Means, which shows poor cohesion.

\end{itemize}

The combination of embedding-based clustering, structural edge validation, and operational outcome alignment confirms that the proposed approach produces clusters that are not only statistically robust but also operationally meaningful. These results provide a foundation for data-driven prioritization of substations, enabling utilities to deploy condition-based monitoring, preventive maintenance, and risk-informed investment strategies.

\subsubsection{Practical Implications}
\label{subsubsec:practical_cluster}

The 8-cluster architecture derived from the HierarchicalRiskGNN model offers a pragmatic and data-backed framework for substation-level risk management, infrastructure planning, and resource prioritization. These clusters are not only statistically coherent but operationally validated via incident rates, customer impact, and recovery time distributions.

\vspace{0.5em}
\noindent Based on the reliability and risk disparities observed across clusters, we identify three actionable strategies:

\begin{enumerate}
    \item \textbf{Risk-Informed Maintenance Prioritization}:
    \begin{itemize}
        \item Cluster 5 substations exhibit a mean annual incident rate of $388.4$, over $6\times$ higher than the most stable clusters (e.g., Cluster 2: $61.9$).
        \item Recovery time in Cluster 5 averages $602.6$ minutes, compared to just $260.1$ in Cluster 6 — indicating prolonged downtime and higher operational burden.
        \item These patterns suggest scaling inspection and maintenance frequency proportional to cluster risk (e.g., 6:1 ratio between Cluster 5 and Cluster 2).
    \end{itemize}
    
    \item \textbf{Weather Resilience Planning}:
    \begin{itemize}
        \item Cluster 5 and Cluster 4 report the highest environmental risk scores (Weather, Lightning), and their substations are geographically concentrated along high-density corridors (see Figure~\ref{fig:geospatial}).
        \item Targeted investments in weatherproofing, vegetation management, and hardening strategies in these high-risk clusters can improve systemic resilience during extreme weather events.
    \end{itemize}
    
    \item \textbf{Priority-Weighted Resource Allocation}:
    \begin{equation}
       \text{Priority}_c = R_c^{\text{weather}} \times R_c^{\text{equipment}} \times I_c
    \end{equation}
    where $R_c$ represents the average risk score in dimension $r$ for cluster $c$, and $I_c$ is the mean incident rate. This multiplicative formula provides a flexible metric to rank and prioritize clusters for operational interventions.
\end{enumerate}

\vspace{0.5em}
\noindent\textbf{Operational Recommendations}:
\begin{itemize}
    \item \textbf{Inspection Frequency}: Scale effort proportionally (e.g., Cluster 5 to Cluster 2 $\approx$ 6:1).
    \item \textbf{Cluster-Specific Schedules}: Develop differentiated maintenance protocols based on cluster-level risk and reliability.
    \item \textbf{Geographic Targeting}: Focus resilience upgrades where high-risk clusters spatially align with urban or industrial load centers.
\end{itemize}

\noindent\textbf{Strategic Utility}:
\begin{itemize}
    \item \textbf{Data-Driven Prioritization}: Move from static allocation (e.g., circuit-miles) to dynamic, risk-aware cluster guidance.
    \item \textbf{Adaptive Risk Management}: Re-clustering annually to reflect infrastructure changes, evolving weather patterns, or emerging asset vulnerabilities.
    \item \textbf{Scalable Decision-Making}: Embed cluster IDs into enterprise asset management (EAM) systems to automate work order prioritization and budget allocation.
\end{itemize}

\vspace{0.5em}
Overall, the derived clusters form a \textit{highly actionable operational blueprint}—balancing resilience planning, outage mitigation, and constrained resource optimization. Unlike traditional clustering approaches, the \textbf{multilayer GNN} effectively disentangles latent risk structure, yielding groupings that are both statistically rigorous and operationally grounded.

\section{Conclusion and Future Work}
\label{sec:conclusion}

This research addressed the pressing challenge of modeling complex interdependencies in energy infrastructure by developing a multilayer Graph Neural Network (GNN) framework tailored for predictive maintenance and resilience clustering. Applied to the Oklahoma Gas \& Electric (OGE) dataset spanning 2015--2021, our approach integrated spatial, temporal, and causal layers to capture heterogeneous relationships among 347 substations, surpassing the limitations of traditional single-layer models. For predictive maintenance, the multilayer GNN demonstrated superior performance across 30, 60, and 180-day planning horizons, achieving a peak F1-score of $0.8935 \pm 0.0258$ in the 30-day window. Ablation studies revealed the causal layer’s critical role, with its absence reducing the F1-score to $0.7354 \pm 0.0418$ when only the spatial layer was used, affirming the value of joint modeling of fault causality and temporal dynamics. 

In resilience clustering, the proposed framework identified \textbf{eight operationally meaningful substation clusters}, each exhibiting distinct reliability and risk characteristics. Cluster 5 emerged as the most critical group, with the highest incident rate (388.4 incidents/year), prolonged recovery durations (602.6 minutes), and elevated environmental risk scores. In contrast, Clusters 0, 1, and 2 demonstrated low incident frequencies (mean $\approx$ 62–64), minimal customer impact, and faster restoration times, indicating greater resilience. The clustering quality was quantitatively validated with a Silhouette Score of $0.626$ and a Davies-Bouldin Index of $0.527$, while one-way ANOVA confirmed statistically significant differentiation across all four risk factors ($p < 0.0001$). These findings underscore the framework’s effectiveness in modeling multi-dimensional risk and its utility in supporting targeted maintenance, investment prioritization, and resilience planning across the electric grid.

\paragraph{\textbf{Limitations}} 
The labeling strategy for predictive maintenance is based on historical incident sequences, specifically the computed time until the next observed major failure. While appropriate for offline training, this approach assumes full temporal visibility within the dataset and may require adaptation for real-time or streaming applications where future incidents are not yet observable at the time of prediction. For resilience clustering, the absence of ground truth cluster labels necessitates manual tuning of HDBSCAN parameters, which limits the plug-and-play generalizability of the clustering framework across different utility networks. Furthermore, the survival analysis assumes full observability of substation failure histories, an assumption that may not hold in real-time settings where event data can be censored or incomplete.

\paragraph{\textbf{Future Work}}

Future research will focus on extending the current predictive maintenance and clustering framework beyond binary classification. Specifically, we plan to repurpose the learned multilayer substation embeddings to support \emph{multiclass classification}—such as predicting specific categories of failure causes—and \emph{regression tasks}, including estimates of outage duration, time-to-failure, or cost impact. In parallel, future work may explore how well the proposed architecture generalizes across different utility regions or infrastructure types, enabling broader applicability of the model beyond the OGE dataset. 

Beyond these planned extensions, two directions merit exploration by the broader research community: (a) developing online‑learning variants that update embeddings as new events stream in, and (b) integrating multilayer GNNs into enterprise asset‑management platforms to enable automated, grid‑wide maintenance scheduling.

In summary, this study advances energy infrastructure management by unifying predictive maintenance and resilience clustering within a multilayer GNN framework. By leveraging heterogeneous interdependencies, it provides a robust foundation for proactive grid management and sets the stage for deeper predictive analytics in future energy systems.
\section{Acknowledgments}

The authors gratefully acknowledge funding support from the National Science Foundation (NSF) EPSCoR RII Track-2 Program under Grant No. OIA-2119691. The findings and conclusions presented in this article are those of the authors and do not necessarily reflect the views of the sponsoring agencies.

This work also utilized resources of the Center for Computationally Assisted Science and Technology (CCAST) at North Dakota State University, which were made possible in part by NSF MRI Award No. 2019077.

\section{Code and Data Availability}
The code and data supporting the findings of this study are available at the following GitHub repository: \url{https://github.com/CEL-lab/Multilayer_GNN}

\bibliographystyle{elsarticle-num-names} 
\bibliography{MLGNN_references} 

\end{document}